\documentclass[conference]{IEEEtran}

\usepackage{graphics, array, url, hyperref, epsfig, amsmath, amssymb, graphicx}
\usepackage{subfigure}
\usepackage{threeparttable}
\usepackage{extarrows}
\usepackage{color}
\usepackage{indentfirst}
\usepackage{float}
\usepackage{cleveref}
\usepackage{fancybox}
\usepackage{multirow}
\usepackage{subfigure}
\usepackage{cite}

\newsavebox{\savepar}
\newenvironment{boxit}{\begin{center} \begin{lrbox}{\savepar}
        \begin{minipage}[b]{80mm}}
        {\end{minipage}\end{lrbox}
      \fbox{\usebox{\savepar}
} \end{center}}

\newenvironment{boxita}{\begin{center} \begin{lrbox}{\savepar}
        \begin{minipage}[b]{135mm}}
        {\end{minipage}\end{lrbox}
      \fbox{\usebox{\savepar}
} \end{center}}

\begin{document}

\title{A Blockchain-based Self-tallying Voting Scheme\\ in Decentralized IoT}

\author{\IEEEauthorblockN{Yannan Li$^{\dagger}$, Willy Susilo$^{\dagger}$, Guomin Yang$^{\dagger}$, Yong Yu$^{\ddagger}$, Dongxi Liu$^{\S}$ and Mohsen Guizani$^{\o}$
}
 \IEEEauthorblockA{$^{\dagger}$Institute of Cybersecurity and Cryptology, School of Computing and Information Technology\\ University of Wollongong, Wollongong, NSW 2522, Australia.\\  Email: yl738@uowmail.edu.au, \{wsusilo, gyang\}@uow.edu.au}
 \IEEEauthorblockA{$^{\ddagger} $ School of Computer Science, Shaanxi Normal University, Xi'an 710062, China.
}
\IEEEauthorblockA{$^{\S}$ Data61, CSIRO, Marsfield, NSW 2122, Australia.
}

\IEEEauthorblockA{$^{\o}$ Dept. of Electrical and Computer Engineering, University of Idaho, Moscow, Idaho, USA.
}
}
%
%
%
%
%
%
\maketitle

\begin{abstract}
The Internet of Things (IoT) is experiencing explosive growth and has gained extensive attention from academia and industry in recent years. Most of the existing IoT infrastructures are centralized, in which the presence of a cloud server is mandatory. However, centralized frameworks suffer from the issues of unscalability and single-point-of-failure. Consequently, decentralized IoT has been proposed by taking advantage of the emerging technology of Blockchain.
Voting systems are widely adopted in IoT, such as a leader election in wireless sensor networks.
Self-tallying voting systems are alternatives to traditional centralized voting systems in decentralized IoT since the traditional ones are not suitable for such scenarios.
Unfortunately, self-tallying voting systems inherently suffer from fairness issues, such as adaptive and abortive issues caused by malicious voters. In this paper, we introduce a framework of self-tallying systems in decentralized IoT based on Blockchain. 
We propose a concrete construction and prove the proposed system satisfies all the security requirements including fairness, dispute-freeness and maximal ballot secrecy. The implementations on mobile phones demonstrate the practicability of our system.

\end{abstract}

\section{Introduction}\label{sec:Introduction}
The Internet of Things (IoT) \cite{Du1}-\cite{Du9} is a system comprising smart devices, actuators, sensors and other objects that connected through the network with the ability to transfer data, share resources and make decisions without man-to-man or man-to-device interaction. Organizations in various industries utilize IoT for better efficiency, convenience and service \footnote{https://internetofthingsagenda.techtarget.com/definition/Internet-of-Things-IoT}. Besides the well-known applications of smart city and smart home, IoT has potential in many other public and private applications, such as manufacturing, agriculture, transportation, and healthcare. In recent years, IoT gains extensive attention and some new extensions of IoT are proposed, such as IIoT and NB-IoT. The number of devices is in an explosive growth. It is estimated that there will be 50 billion IoT devices by the year of 2020 \cite{iot1}.

\emph{Voting systems and IoT.} Voting systems have wide applications in IoT. We list two typical examples here. 1) Leader election in distributed IoT. Leader election is one of the most common and important activities in a distributed IoT, such as wireless sensor networks \cite{decision1}. The goal of leader election is to designate a special node as an organizer to coordinate the tasks in distributed nodes to break the inner symmetry in distributed systems. The peers in the network select and communicate among themselves to vote for the leader.
2) Decision making in IoT systems. One of the most salient features of IoT systems is to collect data and make smarter decisions according to the sensed data. Voting is one of the effective approaches making decisions \footnote{http://healthandlearning.org/wp-content/uploads/2017/04/Decision-Making-Models-Voting-versus-Consensus.pdf}. Devices measure various types of data and leverage diverse methods to analyze the data, which may lead to a different opinion to a specific decision. Then devices vote for a final decision. Take an environmental health IoT as an example, which comprises some smart phones with apps to acquire the environmental parameters with high accuracy including temperature, humidity, noise, and dust. All the parameters are closely related to people's health. Thus, the environmental health IoT is an important reference for healthcare. The smart devices collaborate to make decisions in an environmental health IoT to show whether the current environment is suitable to live or work in.

Nevertheless, most of the IoT implementations are with a centralized infrastructure. Specifically, the devices are linked with the cloud and controlled by a central hub, which suffers from scalability, efficiency and single-point-of-failure issues. To overcome the bottleneck of centralized IoT, the notion of decentralized IoT was proposed \cite{iotarchi}. The devices enjoy the desirable merits of  peer-to-peer (P2P) communication, flexibility and better efficiency to exchange information. A majority of decentralized IoT leverages Blockchain \cite{decentralizediot} to build the underlying P2P network. A San Francisco based startup Helium \footnote{https://www.helium.com/} has built a blockchain machine network for IoT. \footnote{https://internetofbusiness.com/helium-blockchain-machine-network-iot-unleashed/}

Traditional voting systems with a central party to organize the vote and tally the ballots are not suitable for a decentralized IoT framework. As an alternative, self-tallying scheme was proposed, which does not require a third party to tally the ballots and reveal the final result. Instead, after all the voters cast the ballots, anyone can collect the ballots and compute the final results at the same time. However, self-tallying schemes inherently suffer from fairness issues, as the last voter can collect other voters' ballots to compute the final result before casting his/her own ballots. That is he can know the result ahead of schedule. Moreover, if he is not satisfied with the final result, this voter may refuse to reveal his ballot, then other voters are hard to obtain the final result.

\subsection{Our Contributions}

In this paper, we aim to improve the fairness of blockchain-based self-tallying systems for decentralized IoT. The contributions of this article are listed as follows.
\begin{enumerate}
  \item We formalize the system model of self-tallying voting systems based on blockchain in decentralized IoT.
        \item We propose a concrete construction of a blockchain-based self-tallying voting protocol in decentralized IoT, and prove it satisfies fairness, dispute-freeness and maximal ballot secrecy. Specifically, in our construction, we modify the commitment in \cite{MSH17} and recovery phase in \cite{KSRH12} to handle the abortive issues, and suggest that timed commitment can be used to deal with adaptive issues self-tallying voting schemes based on blockchain.
  \item We implement the proposed protocol on a laptop and a mobile phone respectively.
\end{enumerate}
\textbf{Organization.} The rest of the paper is organized as follows. We review the related work and provide some preliminaries in Sec. \ref{sec:relatedwork} and Sec. \ref{sec:Preliminary}, respectively. The security requirements are presented in Sec. \ref{sec:Security}. We build a fair blockchain-based self-tallying voting system for decentralized IoT in Sec. \ref{sec:Construction}. The implementations of the proposed protocol are illustrated in Sec. \ref{sec:Implementation}. Finally, we conclude the paper in Sec. \ref{sec:Conclusion}.


\section{Related work}\label{sec:relatedwork}
\emph{Self-tallying e-voting.} E-voting is a flourishing and fadeless topic in academic research.
In traditional centralized e-voting protocols, a central authority is usually involved for organizing the election and counting the votes. To achieve stronger voter privacy, Kiayias and Yung \cite{KY02} proposed the notion of self-tallying voting, which is a new paradigm in decentralized e-voting systems. In self-tallying systems, tallying is an open procedure in which any party, not only the voters but also the observers, can check the validity of each ballot and perform the computation after collecting all the valid ballots to get the final voting result. They proposed the first concrete construction as well by leveraging a bulletin board, which achieves perfect ballot privacy and dispute-freeness. 
Groth et al.\cite{Groth04} proposed a simpler scheme with better efficiency for each voter. They also constructed an anonymous broadcast channel with perfect message secrecy at the cost of increased round complexity of the protocol, which needs $n+1$ rounds for $n$ voters.
Hao et al. \cite{HRZ10} proposed a self-tallying voting protocol based on a two-round anonymous veto protocol (AV-net). Their protocol provides the same security properties but achieves better efficiency in terms of round complexity. Khader et al. \cite{KSRH12} claimed that \cite{HRZ10} is neither robust nor fair, and they advanced the protocol by adding a commitment phase and a recovery round. However, the recovery phase in their construction ignores the ballots of the abortive voters.

\emph{Blockchain-based e-voting systems.} There are already some existing works on e-voting based on blockchain. The role of blockchain in e-voting protocols varies from schemes to schemes. Most of the works make blockchain as bulletin boards and still employ a trusted authority for voter privacy, such as Follow My Vote \footnote{https://followmyvote.com/}, TIVI \footnote{ https://tivi.io/} \footnote{http://www.smartmatic.com/fileadmin/user\_upload/Whitepaper\_Online\_ Voting\_Challenge\_Considerations\_TIVI.pdf}. Some of the existing works are based on cryptocurrencies, such as Bitcoin \cite{coinvotin}\cite{coinvoting2}. Takabatake et al. \cite{TKO16} proposed a voting protocol based on Zerocoin to enhance voter privacy.
In 2017, McCorry et al. \cite{MSH17} presented Open Vote Network \footnote{https://github.com/stonecoldpat/anonymousvoting} \footnote{https://ethereumfoundation.org/devcon3/sessions/the-open-vote-network-decentralised-internet-voting-as-a-smart-contract/}, the first implementation of a decentralized self-tallying e-voting protocol based on Blockchain. The commitment in \cite{MSH17} is the hash of the vote, which is irrecoverable if a voters refuses to cast his ballot in the voting phase.

\section{Preliminaries}\label{sec:Preliminary}
In this section, we provide some preliminaries used in our construction.

\subsection{Intractable Assumptions}
1) \textbf{\emph{Discrete Logarithm (DL) Assumption.}}

\vspace{3pt}

Let $\lambda$ be a security parameter and $G=<g>$ denotes a cyclic group of prime order $p$. DL problem \cite{JKatzSign} is that, given a tuple $(g,g^a)\in G$ and output $a\in Z_p$, where $Z_p$ is the set of non-negative integers smaller than $p$. DL assumption holds if for any polynomial-time algorithm $\mathcal{A}$, the following advantage $\textbf{Adv}_\mathcal{A}^{\textbf{DL}}$ is negligible in $\lambda$. $$\textbf{Adv}_\mathcal{A}^{\textbf{DL}}(\lambda)=Pr\Big[\mathcal{A}(g,g^a)\rightarrow a\Big]$$


%
%
2) \textbf{\emph{Decisional Diffie-Hellmam (DDH) Assumption}}

\vspace{3pt}

Let $\lambda$ be a security parameter and $G=<g>$ denotes a cycle group of prime order $p$. DDH problem \cite{JKatzSign} states that given a tuple $(g,g^a,g^b,$ $g^{(1-x)ab+xc})\in G$ and output $x\in \{0,1\}$. DDH assumption holds if for any polynomial-time algorithm $\mathcal{C}$, the following advantage $\textbf{Adv}_\mathcal{C}^{\textbf{DDH}}(\lambda)$ is negligible in $\lambda$.
\small{ $$\textbf{Adv}_\mathcal{C}^{\textbf{DDH}}(\lambda)=\Big|\!Pr\Big[\mathcal{C}(g,g^a,g^b,g^{ab})=\!\!1\Big]-Pr\Big[\mathcal{C}(g,g^a,g^b,g^{c})=\!1\!\Big]\!\Big|$$}

\subsection{ElGamal encryption}

%
%

ElGamal encryption \cite{elgamal} is semantically secure under the DDH assumption. Another merit of ElGamal encryption is its inherent homomorphism. The ciphertexts of $m_0,m_1$ can be easily aggregated to obtain the ciphertext of $m_0m_1$. A distributed ElGamal cryptosystem \cite{Brandt05} is a generalization of ElGamal encryption, which contains the following algorithms.

{\sf{Setup}}. Suppose there are $n$ users in the system, and the key pairs of the $i$-th user are $(x_i,y_i=g^{x_i})$. Each user publishes his public key, and the common public key can be generated in a distributed manner\cite{Ped91} as $y=\prod\limits_{i=1}^n {y_i}$.

{\sf{Enc}}. To encrypt a message $m$, randomly choose $r$, and compute a ciphertext $(c_1,c_2)$ of $m$ as $(g^r,y^r\cdot g^m)$.

{\sf{Dec}}. Each user computes and broadcasts the partial decryption key $c_1^{x_i}$. Then the decryption can be done by computing $$g^m=c_2/{\prod_{i=1}^n{c_1}^{x_i}}=c_2/{{c_1}^{x_1+\cdots+x_n}}$$.

\subsection{Commitment}
A commitment scheme allows a user to commit to a selected statement, which is hidden to others during the \emph{Commit} phase, but can be revealed by the user in the \emph{Open} phase. A commitment scheme contains the following two properties:

\begin{itemize}
  \item \emph{Binding}. The committer cannot change the statement after he has committed to the statement.
  \item \emph{Hiding}. The receiver knows nothing about the committed statement before the committer opens the commitment.
\end{itemize}

\subsection{Zero-Knowledge Proof of knowledge and $\Sigma$-Protocol}
Let $R={(x,w)}$ be a binary relation, where $x$ is the common input and $w$ is a witness. A zero-knowledge proof of knowledge is a protocol in which a prover $\mathcal{P}$ proves to a verifier $\mathcal{V}$ that it knows a witness $w$ for which $(x,w)\in R$ without revealing anything.

\begin{figure}[htbp]
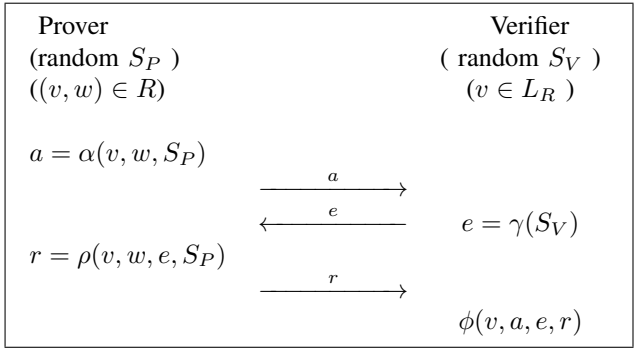

\begin{boxit}
\begin{center}
\begin{tabular}{llcl}
{ \centering{Prover} }            &       &   { \centering{ Verifier} }\\
(\centering{random $S_P$} ) &       &  ( {random $S_V$ })\\
($(v,w)\in R$)&       &   ({$v\in L_R$ })\\
            &   &\\
  $a=\alpha(v,w,S_P)$  && \\
    & $\stackrel{a}{\overrightarrow{\hspace{2 cm}}}$          &  \\
    & $\stackrel{e}{\overleftarrow{\hspace{2 cm}}}$          & $e=\gamma(S_V)$ \\
  $r=\rho(v,w,e,S_P)$ && \\
  & $\stackrel{r}{\overrightarrow{\hspace{2 cm}}}$          &  \\
   && $\phi(v,a,e,r)$\\
\end{tabular}
\end{center}
\end{boxit}
\begin{center}
\caption{$\Sigma$ Protocol}\label{ZKP0}
\end{center}
\end{figure}

A $\Sigma$-Protocol is a way to obtain efficient zero-knowledge proof. A protocol is a $\Sigma$-Protocol for relation $R$ if it has 3-move as shown in Fig \ref{ZKP0}. 1) $\mathcal{P}$ sends a message $a$ to $\mathcal{V}$. 2) $\mathcal{V}$ sends a random $t$-bit challenge $e$ to $\mathcal{P}$. 3) $\mathcal{P}$ sends a response $r$, and $\mathcal{V}$ decides to accept or reject based on the verification algorithm.

A $\Sigma$-protocol has the following properties.
\begin{itemize}
  \item Completeness. If $\mathcal{P}$ and $\mathcal{V}$ follow the protocol on input $x$ and $w$, where $(x,w)\in R$, the verifier always accepts the prover's proof.
  \item Special soundness. For any $x$ and any accepting conversations on $x$ with the same message $a$ and different challenges $(a,e,r)$ and $(a,e',r')$, where $e\neq e'$, one can efficiently extract $w$ such that $(x,w)\in R$.
  \item Honest verifier zero-knowledge(HVZK). There is a polynomial-time simulator, which on input $x$ and a challenge $e$ outputs an accepting conversation with the form $(a,e,r)$, with the same probability distribution as conversations between the honest $\mathcal{P}$ and $\mathcal{V}$ on input $x$.
\end{itemize}

The special soundness property implies that the error probability of this proof system is always $2^{-t}$.

\begin{figure}[h]
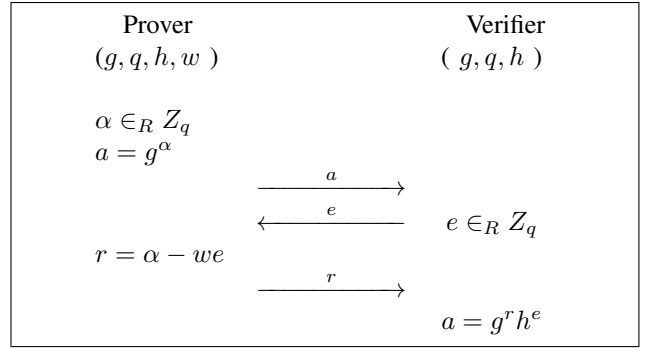

\begin{boxit}
\begin{center}
\begin{tabular}{llcl}
{ ~~Prover }            &       &   { ~~Verifier }\\
($g,q,h,w$ ) &       &  ( $g,q,h$ )\\
            &   &\\
  $\alpha\in_R Z_q$    &  &  \\
  $a={g}^{\alpha}$  && \\
    & $\stackrel{a}{\overrightarrow{\hspace{2 cm}}}$          &  \\
    & $\stackrel{e}{\overleftarrow{\hspace{2 cm}}}$          & $e\in_R Z_q$ \\
  $r=\alpha-we$ && \\
  & $\stackrel{r}{\overrightarrow{\hspace{2 cm}}}$          &  \\
   && $a=g^{r}{h}^{e}$\\
\end{tabular}
\end{center}
\end{boxit}
\begin{center}
\caption{$\Sigma$ protocol for a DL tuple}\label{Schnorr}
\end{center}
\end{figure}

Now we show a special case of the definition above by Schnorr to prove a DL relation (Fig. \ref{Schnorr})\cite{Schnorr91}. Let $G$ be a group of order $q$ with a generator $g$. $\mathcal{P}$ and $\mathcal{V}$ have common inputs $(g,q, h\in G)$, $\mathcal{P}$ has a private input $w$ such that $g^w =h$.

$\Sigma$-Protocol is efficient to prove AND, OR and arbitrary combination of AND/OR statement. More details can be found in \cite{CDS94}\cite{DDPY94}\cite{Cam97}.


\section{System and Security Model}\label{sec:Security}
In this section, we propose the system model of blockchain-based self-tallying voting system for decentralized IoT and list the necessary security requirements and security model of a self-tallying voting protocol.
\vspace{4 pt}

\subsection{System model}
\begin{figure}
  \centering
  \includegraphics[width=0.5\textwidth]{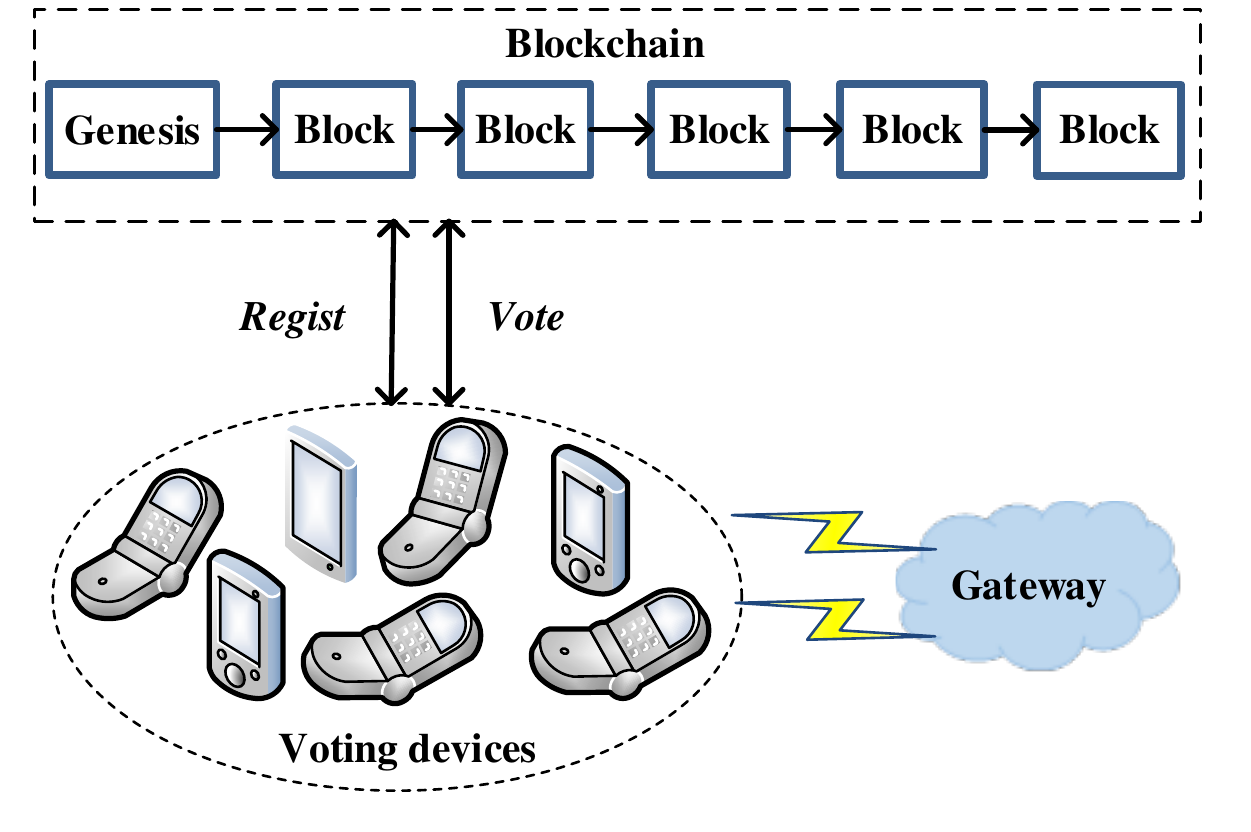}\\
  \caption{ The framework of blockchain-based self-tallying voting system in a decentralized IoT}\label{IoT}
\end{figure}

The framework of a blockchain-based self-tallying voting protocol for decentralized IoT system is shown in Fig. \ref{IoT}. There are three roles in the sytem, smart devices, gateway and a blockchain. The IoT system is equipped with many smart devices, which are regarded as voting devices. A blockchain is leveraged to achieve a P2P overlay network and can also fulfill device management \cite{devicemanage} and a bulletin board. Each device needs to register when they first enroll in the system and cast the ballots through gateway to blockahin. After collecting the ballots from the blockchain, the results can be obtained immediately to make decisions for the whole IoT system.
\subsection{System components}

A blockchain-based self-tallying voting system in a decentralized IoT consists of the following algorithms. Suppose there are $n$ voting devices in the system denoted as voter $\overline{V_i}$.

\textbf{Setup($(k,n)\rightarrow(sk_i,pk_i)$).} This is a probabilistic algorithm that takes a security parameter $k$ and the number of voters $n$ as input and outputs the private and public key pair $(sk_i,pk_i)$ for each voter $\overline{V_i}$.

\textbf{Commit($(v_i,\{pk_j\}_{(j\neq i)})\rightarrow C_i$).} This algorithm is run by each voter $\overline{V_i}$. On input a vote $v_i$ and other voter $\overline{V_j}$'s public key $\{pk_j\}_{(j\neq i)}$, it outputs a commitment $C_i$ and a corresponding zero-knowledge proof, and publishes $C_i$ and the proof on the blockchain.

\textbf{Vote($(v_i,sk_i,\{pk_j\}_{(j\neq i)})\rightarrow V_i$).} This algorithm is run by each voter $\overline{V_i}$. On input a vote $v_i$, the private key $sk_i$, and the other voter $\overline{V_j}$'s public key $\{pk_j\}(j\neq i)$, it outputs a ballot $V_i$ and a zero-knowledge proof to prove the ballot is in the right form, and publishes $V_i$ and the proof on the blockchain.

\textbf{Tally($\{V_i\}_{(i\in n)}\rightarrow Result$).} This is a deterministic algorithm that takes all the ballots $\{V_i\}(i\in n)$ as input, it outputs the election result.

\textbf{Recover($(\{sk_i\}_{(i\in n)},C_i)\rightarrow v_i$).} This algorithm is to recover the abortive voter's vote. On input the abortive voter's commitment $C_i$ and all the other voters' private key $\{sk_j\}_{(j\neq i)}$, it outputs the abortive voter's vote $v_i$.

\subsection{Security requirements}
A self-tallying protocol is supposed to satisfy the following security requirements.
\begin{itemize}
  \item \textbf{Maximal ballot secrecy}. A partial tally of the ballots can be accessed only by a collusion of all remaining voters.
  \vspace{2 pt}
  \item \textbf{Self-tallying}. After all the voters cast their ballots, anyone is able to compute the voting results with all the ballots.
  \vspace{2 pt}
  \item \textbf{Fairness}. Freeness means that nobody has the priority to get a partial tally ahead of schedule. Self-tallying protocols always suffer from fairness issues, including abortive issues and adaptive issues. Abortive issues indicate that some of the users refuse to reveal their votes and abort before casting their ballots, then the final results won't be revealed. Adaptive issues state that the last voter has the priority to get the final results first and thus may affect his choice and even gives up in this vote, thus will cause an abortive issue.
  \vspace{2 pt}
  \item \textbf{Dispute-freeness}. This property states that anybody can check whether the voters follow the protocol or not. This is an extension of universal verifiability.
\end{itemize}

\subsection{Security model}\label{subsec:Securitymodel}
In this section, we formalize the security model for \emph{maximal ballot secrecy}.

Suppose there are maximal $n-2$ corrupted voters in the \emph{maximal ballot secrecy} game, who are fully controlled by the adversary. The adversary can make queries to the commitments as well as the corrupted users' ballots, and also get access to the final result of the election.
And later in the challenge phase, given two ballots from different votes $\{0,1\}$ for the two uncorrupted voter, the adversary needs to tell which of the two ballots is from the vote 1.
The detailed security model is as follows.

\textbf{Maximal ballot secrecy (MBS):} We say a self-tallying voting scheme is MBS-secure, if no polynomial bounded adversary $\mathcal{A}$ has a non-negligible advantage against a challenger $\mathcal{C}$ in the following game.

\textbf{Initial.} There are $n$ voters in the game. $\mathcal{A}$ declares two target voters $V_s,V_t$ to be challenged upon. The other voters are regarded as corrupted users. $\mathcal{C}$ randomly chooses one from $\{V_s,V_t\}$ and denotes as $v^*$. Set the vote of $\overline{V}^*$ as 1, and the other one as 0.

\textbf{Setup.} $\mathcal{C}$ generates the private and public key pairs for each voter. Then $\mathcal{C}$ forwards all the public keys and the corrupted users' private keys to $\mathcal{A}$.

\textbf{Queries.} $\mathcal{A}$ can choose any ballots for the corrupted users and make some queries including the Commit queries and the Vote queries corresponding to the chosen ballots.
  \begin{itemize}
    \item {Commit queries.} $\mathcal{A}$ can query the commitment for a vote. Then $\mathcal{C}$ generates the commitment and records the ballot and the commitment in the list $\mathcal{L_C}$.
    \item {Vote queries.} $\mathcal{A}$ can make queries on the votes generated by any user other than $V_s,V_t$.
  \end{itemize}

\textbf{Challenge.} $\mathcal{C}$ outputs two challenge ballots on behalf of the uncorrupted voters $V_s$ and $V_t$ chosen in the \emph{Initial} phase.

\textbf{Tally.} $\mathcal{A}$ can compute the final result of the election according to the collected ballots.

\textbf{Guess.} $\mathcal{A}$ outputs a guess $guess$ to determine which one between $V_s$ and $V_t$ has cast the ballot of 1.

\vspace{4 pt}


In the above model, the reason we set two challenge votes rather than one is to prevent the adversary deducing the challenged vote from his known information. Specifically, as the adversary can control the ballot of the corrupted voters and will obtain the election result after collecting the challenge vote, if there is only one challenge vote, the adversary can have a non-negligible advantage in the guessing game. After collecting the votes together, the adversary can do the tallying by itself to know the election result. To see why we set the different ballots for the two challenge votes, let's suppose the following situation. If we set the challenge vote with the same ballot from $\{0,1\}$, and all the corrupted voters controlled by the adversary vote the same ballot, then the adversary can get the knowledge about the challenge vote easily after knowing the results, in which the advantage $\epsilon$ is non-negligible.

\vspace{4 pt}

\textbf{Definition 1.} The voting scheme is MBS secure if for any polynomial time adversary,
$$|Pr[guess=\overline{V}^*]-1/2|\leq \epsilon,$$ where $\epsilon$ is negligible.

\section{Construction}\label{sec:Construction}
\begin{figure}[h]
  \centering
  \includegraphics[width=0.5\textwidth]{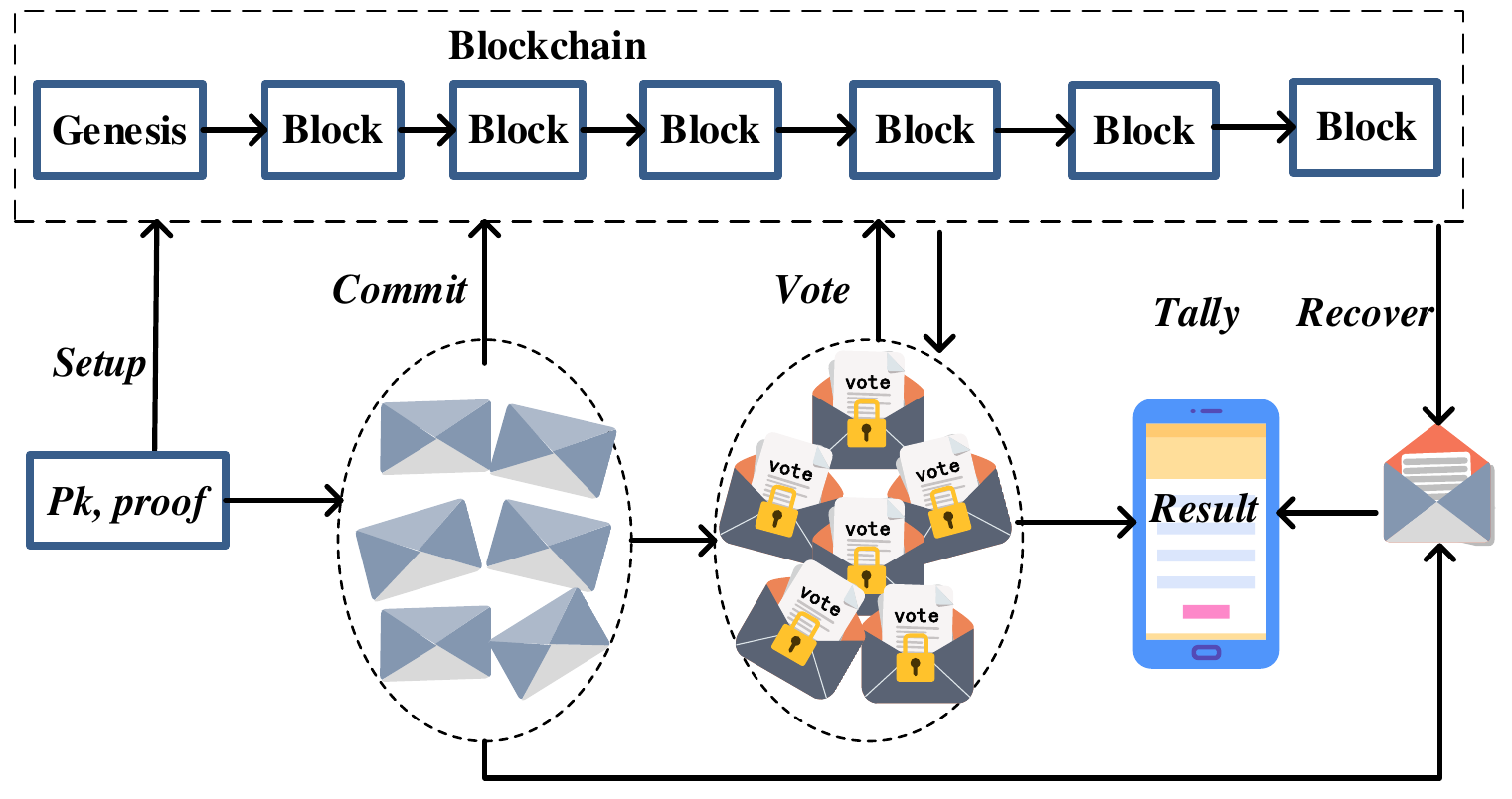}\\
  \caption{The workflow of blockchain-based self-tallying voting system. }\label{workflow}
\end{figure}

In this section, we present a concrete construction of self-tallying voting system with the help of blockchain.
As shown in Fig. \ref{workflow}, the system contains three phases, the {\sf{Pre-vote}} phase, which includes \emph{Setup} and \emph{Commit} algorithms, {\sf{Vote}} phase, which includes \emph{Vote} algorithm, and {\sf{After-vote}} phase, which includes \emph{Tally} and \emph{Recover} algorithms. In the {\sf{Pre-vote}} phase, the system is initialized and the voters register to obtain the private-public key pairs. Voters put their public keys together with the zero-knowledge proof for the corresponding private keys on the blockchain. {\emph{Commit}} is to ensure fairness. If voters skip commit part and cast their ballots directly, the last voter has the priority to access the final result ahead of schedule. 
In this phase, other voters cannot see the vote but only the commitment of the vote, thus, the voters need a zero-knowledge proof to prove the committed vote is in the right form. Later if the last voter refuses to vote, other voters can recover the ballot according to the commitment and get the result. In {\sf{Vote}} phase, the voters cast their encrypted ballots. In {\sf{After-vote}} phase, by collecting all the ballots from the blockchain, the final result can be obtained publicly by anyone. {\emph{Recover}} is an optional phase which can be called when the last voter does not follow the rule to cast his/her ballot in the sense that the ballot can be recovered with the corresponding commitment and the help of all the other voters.

\subsection{Dealing with abortive issues}

\textbf{Basic idea.} The existing approaches to deal with abortive issue are adding a recovery phase, in which the abortive users are excluded by removing their ballots, and tally the ballots from the remaining voters. However, an abort may be caused by some user knowing the unwanted result and against revealing the result. So, simply removing the votes will bring a different voting result. Thus, we modified the recovery phase in \cite{MSH17}. In our modification, if the last voter quits after making a commitment, then his/her ballot can be revealed according to the corresponding commitment with the cooperation of all the other voters. The detailed construction is as follows.

\vspace{5pt}



\begin{figure}[h]
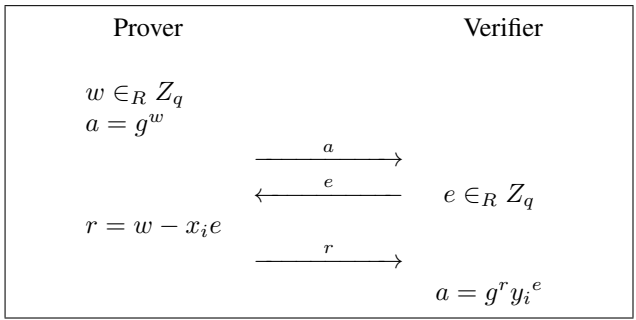

\begin{boxit}
\begin{center}
\begin{tabular}{llcl}
{ ~~Prover }            &       &   { ~~Verifier }\\
            &   &\\
  $w\in_R Z_q$    &  &  \\
  $a={g}^{w}$  && \\
    & $\stackrel{a}{\overrightarrow{\hspace{2 cm}}}$          &  \\
    & $\stackrel{e}{\overleftarrow{\hspace{2 cm}}}$          & $e\in_R Z_q$ \\
  $r=w-x_ie$ && \\
  & $\stackrel{r}{\overrightarrow{\hspace{2 cm}}}$          &  \\
   && $a={g}^{r}{y_i}^{e}$\\
\end{tabular}
\end{center}
\end{boxit}
\begin{center}
\caption{Zero-knowledge proof for Setup}\label{ZKP1}
\end{center}
\end{figure}

\textbf{Setup($(k,n)\rightarrow(x_i,y_i)$).} On input a security parameter $k$, and the number of voters $n$, it initializes the system. Each voter $\overline{V_i}$ chooses a random private key $x_i\in Z_q^*$, and computes the public key $g^{x_i}$. Then $\overline{V_i}$ generates a zero-knowledge proof as ZKPoK$_1\{(x_i):y_i=g^{x_i}\}$(cf. Fig. \ref{ZKP1}). The public key and the corresponding ZKP are published to Blockchain.

\begin{figure*}[htbp]
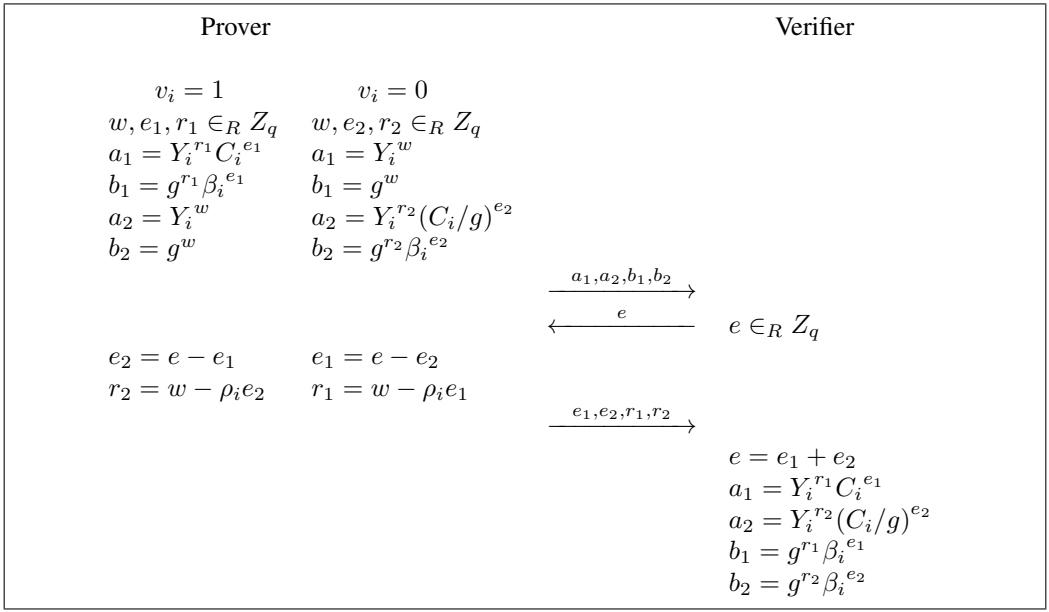

\begin{boxita}
\begin{center}
\begin{tabular}{llcl}
{ ~~~~~~~~~Prover }            &   &    &   { ~~~ Verifier }\\
            &  & &\\
  {~~~~~$v_i=1$}  &  $~~~~~v_i=0$  &   &  \\
  $w,e_1,r_1\in_R Z_q$  & $w,e_2,r_2 \in_R Z_q$    &  &  \\
  $a_1={Y_i}^{r_1}{C_i}^{e_1}$ &$a_1={Y_i}^{w}$ && \\
  $b_1={g}^{r_1}{\beta_i}^{e_1}$ &$b_1=g^w $ & &  \\
  $a_2={Y_i}^w$  &$a_2={Y_i}^{r_2}{(C_i/g)}^{e_2}$ &  &  \\
  $b_2=g^w$ &$ b_2={g}^{r_2}{\beta_i}^{e_2}$ & &  \\
    && $\stackrel{a_1,a_2,b_1,b_2}{\overrightarrow{\hspace{2 cm}}}$          &  \\
    && $\stackrel{e}{\overleftarrow{\hspace{2 cm}}}$          & $e\in_R Z_q$ \\
  $e_2=e-e_1 $&$e_1=e-e_2$&& \\
  $r_2=w-\rho_ie_2$&$r_1=w-\rho_ie_1$ && \\
  && $\stackrel{e_1,e_2,r_1,r_2}{\overrightarrow{\hspace{2 cm}}}$          &  \\
   & & &  $e=e_1+e_2$\\
   &&& $a_1={Y_i}^{r_1}{C_i}^{e_1}$\\
   &&& $a_2={Y_i}^{r_2}{(C_i/g)}^{e_2}$\\
   &&& $b_1={g}^{r_1}{\beta_i}^{e_1}$\\
   &&& $b_2={g}^{r_2}{\beta_i}^{e_2}$\\
\end{tabular}
\end{center}
\end{boxita}
\begin{center}
\caption{Zero-knowledge proof of knowledge for {\sf{Commit}}}\label{ZKP2}
\end{center}
\end{figure*}

\textbf{Commit($v_i,\{y_j\}\rightarrow C_i$).} Before casting a ballot, each voter $\overline{V_i}$ collects the other voters' public key ${y_j}_{(j\neq i)}$. To generate a commitment to the vote, $\overline{V_i}$ chooses a random $\rho_i$ and publishes $\beta_i=g^{\rho_i}$. $\overline{V_i}$ makes the commitment $C_i=g^{v_i}{Y_i}^{\rho_i}$ to ensure fairness, where $v_i$ is the vote from \{0,1\} and $Y_i=\prod_{j=1,j\neq i}^{n}y_j$. The voters also need to generate a zero-knowledge proof to prove that the commitment is in the right form (cf. Fig. \ref{ZKP2}) as
$$ZKPoK_2\{(\rho_i):(C_i={Y_i}^{\rho_i}\vee C_i=g\cdot{Y_i}^{\rho_i})\wedge \beta_i=g^{\rho_i}\}.$$ And then put the commitment and zero-knowledge proof on the Blockchain.

\begin{figure*}[htbp]
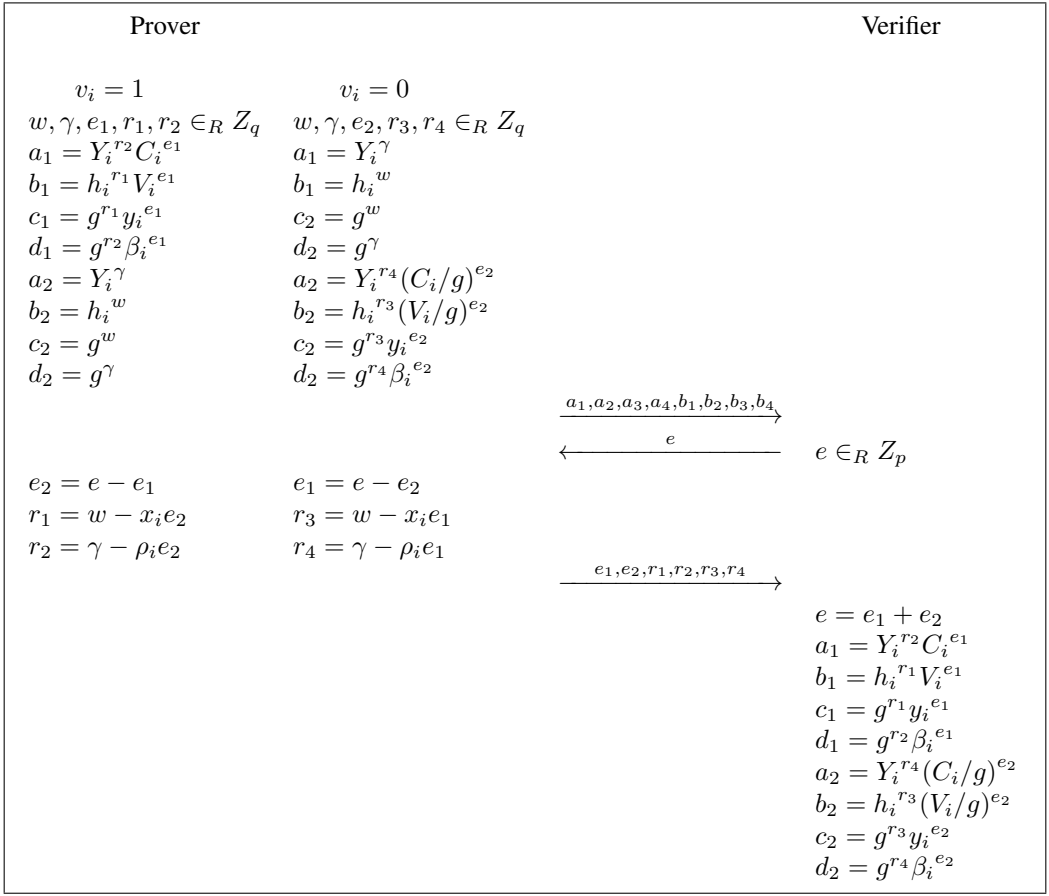

\begin{boxita}
\begin{center}
\begin{tabular}{llcl}
{ ~~~~~~~~~~Prover }            &   &    &   { ~~~ Verifier }\\
            &  & &\\
  {~~~~~$v_i=1$}  &  $~~~~~v_i=0$  &   &  \\
  $w,\gamma,e_1,r_1,r_2\in_R Z_q$  & $w,\gamma,e_2,r_3,r_4 \in_R Z_q$    &  &  \\
  $a_1={Y_i}^{r_2}{C_i}^{e_1}$ &$a_1={Y_i}^{\gamma}$ && \\
  $b_1={h_i}^{r_1}{V_i}^{e_1}$ &$b_1={h_i}^w $ & &  \\
  $c_1={g}^{r_1}{y_i}^{e_1}$ &$c_2=g^w $ & &  \\
  $d_1={g}^{r_2}{\beta_i}^{e_1}$ &$d_2=g^\gamma $ & &  \\
  $a_2={Y_i}^\gamma$  &$a_2={Y_i}^{r_4}{(C_i/g)}^{e_2}$ &  &  \\
  $b_2={h_i}^w$ &$ b_2={h_i}^{r_3}({V_i/g})^{e_2}$ & &  \\
  $c_2=g^w$ &$ c_2={g}^{r_3}{y_i}^{e_2}$ & &  \\
  $d_2=g^\gamma$ &$ d_2={g}^{r_4}{\beta_i}^{e_2}$ & &  \\
    && $\stackrel{a_1,a_2,a_3,a_4,b_1,b_2,b_3,b_4}{\overrightarrow{\hspace{3 cm}}}$          &  \\
    && $\stackrel{e}{\overleftarrow{\hspace{3 cm}}}$          & $e\in_R Z_p$ \\
  $e_2=e-e_1 $&$e_1=e-e_2$&& \\
  $r_1=w-x_ie_2$&$r_3=w-x_ie_1$ && \\
   $r_2=\gamma-\rho_i e_2$&$r_4=\gamma-\rho_i e_1$ && \\
  && $\stackrel{e_1,e_2,r_1,r_2,r_3,r_4}{\overrightarrow{\hspace{3 cm}}}$          &  \\
   & & &  $e=e_1+e_2$\\
   &&& $a_1={Y_i}^{r_2}{C_i}^{e_1}$\\
   &&& $b_1={h_i}^{r_1}{V_i}^{e_1}$\\
   &&& $c_1={g}^{r_1}{y_i}^{e_1}$\\
   &&& $d_1={g}^{r_2}{\beta_i}^{e_1}$\\
  &&& $a_2={Y_i}^{r_4}{(C_i/g)}^{e_2}$\\
   &&& $b_2={h_i}^{r_3}({V_i/g})^{e_2}$\\
   &&& $c_2={g}^{r_3}{y_i}^{e_2}$\\
   &&& $d_2={g}^{r_4}{\beta_i}^{e_2}$ \\
\end{tabular}
\end{center}
\end{boxita}
\begin{center}
\caption{Zero-knowledge proof  of knowledge for {\sf{Vote}}}\label{ZKP3}
\end{center}
\end{figure*}

\textbf{Vote($v_i,x_i,\{y_j\}\rightarrow V_i$).} To ensure the secrecy of the vote, all voters encrypted their votes as $V_i={h_i}^{x_i}g^{v_i}$, where $h_i=\prod_{j=1}^{i-1}g^{x_j}/\prod_{j=i+1}^{n}g^{x_j}$. And generate a zero-knowledge proof to prove that the vote $v_i$ is the same as the one in the commitment. The statement (cf. Fig. \ref{ZKP3}) is
\small{\begin{align*}
ZKPoK_3\{(x_i,\rho_i)\!:\!(C_i={Y_i}^{\rho_i}\!\wedge\! V_i={h_i}^{x_i}\wedge y_i=g^{x_i}\!\wedge\!\beta_i=g^{\rho_i})\\
\vee(C_i=g\cdot{Y_i}^{\rho_i}\wedge V_i=g\cdot{h_i}^{x_i}\wedge y_i=g^{x_i}\wedge\beta_i=g^{\rho_i})\}.
\end{align*}}
Then publish the ballot on the Blockchin.

\vspace{4pt}

\textbf{Tally($\{V_i\}\rightarrow result$).} To tally the votes, one can collect all the ballots and compute $\prod_{i=1}^{n}{h_i}^{x_i}g^{v_i}=g^{\sum_{i=1}^nv_i}$. $\sum_{i=1}^nv_i$ is a small set that can be easily obtained.

\vspace{4pt}

\textbf{Recover($(C_i,\{x_i\})\rightarrow v_i$).} If the last voter $\overline{V_i}$ does not cast his ballot in Vote phase, then each of the remaining voters $V_j(1\leq j\leq n,j\neq i)$ publish a recover factor for $\overline{V_i}$ as $R_{ij}={y_j}^{\rho_i}={\beta_i}^{x_j}$ together with a zero-knowledge proof to prove that it is in the right form (cf. Fig. \ref{ZKP1}). The value of $g^{v_i}$ can be computed as $g^{v_i}=C_i/\prod_{j=1,j\neq i}^{n}{y_j}^{\rho_i}$. Then the value of $v_i$ is easy to get as there are only two candidates.

To compute the final result of the election, each remaining voter $\overline{V_j}$ publishes $\hat{h_j}^{x_j}$, where $\hat{h_j}=\prod_{k=j+1,k\neq i}^ny_j/\prod_{k=1,k\neq i}^{j-1}y_j$ and a ZKPoK to prove the knowledge of $x_j$ as in Fig. \ref{ZKP1}. Now everyone can compute the result of the remaining voters as $g^{\sum_{j\neq i}v_j}=\prod_{j\neq i}\hat{h_j}^{x_j}V_j.$ So the final result of this election is $\sum_{j\neq i}v_j+v_i$.

%

We note that the proofs in Fig. \ref{ZKP2} and Fig. \ref{ZKP3} are three-move interactive protocols with the techniques in \cite{CDS94}, which can be transformed into non-interactive protocols following Fiat-Shamir's heuristics \cite{FS86} by setting $e$ be a hash value of a secure hash function.

\subsection{Dealing with adaptive issues}
The adaptive issues seem inevitable in self-tallying protocols from its definition because the last voter holding the ballot has the priority to access the final results ahead of the other voters. We suggest to use time-locked primitives \cite{RSW96}\cite{encoding} to deal with the adaptive issues in voting systems. Time-lock encryption allows users to get the results only after a certain time \cite{Jager15}\cite{LJ18}. Once the deadline is past, the decryption can be performed immediately. It is stated in \cite{Jager15} that time-locked encryption can be achieved by using a witness encryption with Blockchain as the computational reference clock. We borrow this idea in our proposed scheme by encrypting the vote with a witness encryption and the witness can be produced by Blockchain after certain time. And the blockchain can also be the computational reference clock to measure the ``certain" time, say after generating several blocks. Then the votes can be decrypted once the deadline is past and thus all the voters and observers can do the tallying to get the result simultaneously.

\section{Security Proof}\label{sec:Analysis}
In this section, we show the proposed protocol satisfies all the security requirements presented in Sec. \ref{sec:Security}.

\vspace{4 pt}

Firstly, we show the zero-knowledge proof of knowledge in Fig. \ref{ZKP1}, Fig. \ref{ZKP2} and Fig. \ref{ZKP3} satisfy completeness, special soundness \cite{CDS94,DDPY94} and honest verifier zero-knowledge (HVZK). We show the detailed proof of Fig. \ref{ZKP2} and omit the proofs for other Figures, as the proofs are quite similar.

\vspace{4 pt}
\textbf{Theorem 1.} The proof in Fig. \ref{ZKP2} satisfies completeness, special soundness and honest verifier zero-knowledge.

\vspace{2 pt}

\emph{Proof.} We omit the proof for completeness as it's straightforward to verify.

The witness for the statement in $ZKP_2$ is $\rho_i$. To prove special soundness, the goal is to extract a witness from the three-move interaction with two accepting conversations in polynomial time.
Given the two accepting conversations with the same values in the first round, different random numbers in the second round and different responses in the third round as $(a_1,a_2,b_1,b_2,e,e_1,e_2,r_1,r_2)$ and $(a_1,a_2,b_1,b_2,e,e_1',e_2',r_1',r_2')$, it can be checked easily that one of the following holds $\rho_i'=(r_1-r_1')/(e_1-e_1')$ or $\rho_i'=(r_2-r_2')/(e_2-e_2')$.

To prove HVZK, there exists a simulator $\mathcal{S}$, who is given a random $e$, it randomly chooses $r_1,r_2,e_1,e_2$, where $e=e_1+e_2$, and computes the conversation as $({Y_i}^{r_1}{C_i}^{e_1},$
${Y_i}^{r_2}{C_i/g}^{e_2},{g}^{r_1}{\beta_i}^{e_1},{g}^{r_2}{\beta_i}^{e_2},e,,e_1,e_2,r_1,r_2)$, which is an accepting conversation. It is indistinguishable from the one generated by the honest prover.

\vspace{4 pt}

Next, we prove the proposed scheme is MBS secure if ZKPoK is zero-knowledge and the DDH assumption holds.

\textbf{Theorem 2.} If there exists an adversary that can win the guess game in MBS security model with a non-negligible advantage, then we can build an algorithm $\mathcal{B}$ that can break the zero-knowledge of the ZKPoK and the DDH problem.

\vspace{4 pt}

\emph{Proof.} Suppose there are $n$ voters $\overline{V_1},\cdots,\overline{V_n}$ in the game. The challenger $\mathcal{C}$ can interact with the adversary $\mathcal{A}$.
We list a sequence of games \cite{games} to prove Theorem 2. We denote Pr[Win$_i$] as the winning probability of an adversary (guessing correctly) in Game$_i$.

\vspace{4 pt}

\textbf{Game 0:} This is the original Game defined in section \ref{subsec:Securitymodel}. $\mathcal{A}$ chooses two target voters $\overline{V_s},\overline{V_t}$ to challenge upon and forwards them to $\mathcal{C}$. $\mathcal{C}$ tosses a coin to decide that one of the voters from $\{\overline{V_s},\overline{V_t}\}$ votes 1 and the other one votes 0. The reason that we do in this way is to let $\mathcal{A}$ knows nothing even from the tally result. The one who votes 1 is denoted by $\overline{V}^*$. The challenges are denoted as $\{C_s^*,\pi_s^*,V_s^*\}$ and $\{C_t^*,\pi_t^*,V_t^*\}$, where $C_k^*$ is the commitment of the vote, $\pi_k^*$ represents all the ZKPoK in the scheme, $V_k^*$ is the ballot, $k \in \{s,t\}$. The adversary outputs a guess $guess$, then from the definition of the MBS game, we have $$\Pr[{\sf{Win}}_0]=\Pr[guess=\overline{V}^*].$$

\textbf{Game 1.} Game 1 is the same as Game 0 with one difference. $\mathcal{C}$ runs a simulator $\mathcal{S}$ as in \textbf{Theorem 1.}, and replaces all the zero-knowledge proofs $(\pi_s^*,\pi_t^*)$ with the simulated proofs $(\pi,\pi')$ without using the real witness. The setting is indistinguishable from $\mathcal{A}$'s view.
If $\mathcal{A}$ can distinguish between the two settings in \textbf{Game 0} and \textbf{Game 1} with a non-negligible advantage, then we can use the adversary to construct an algorithm $\mathcal{B}$ to violate {\sf{Zero-Knowledge}} of ZKPoK. Thus, the adversary's winning probability in \textbf{Game 1} satisfies the following equation. $$|\Pr[{\sf{Win}}_1]-\Pr[{\sf{Win}}_0]|\leq \epsilon_{ZK}$$

\textbf{Game 2.} Game 2 is the same as Game 1 with one difference. $\mathcal{C}$ replaces the commitment $C_s^*$ with a random number $C_s$. The two settings are indistinguishable from $\mathcal{A}$'s view.
Specifically, $\mathcal{C}$ generates private and public key pairs for the voters other than $\{\overline{V_s},\overline{V_t}\}$. Then set the public key for $\overline{V_t}$ as $g^a$, $\beta_s$ as $g^b$, $R\in \{g^{ab},g^r\}$, where $r$ is a random number. $\mathcal{C}$ sets $C_s'=g^{v_s}R\cdot{(g^b)}^{\sum_{i=1,i\neq s,t}^nx_i}$
Clearly, if there is a difference in the adversary's winning probability between \textbf{Game 1} and \textbf{Game 2}, we can use the adversary to construct an algorithm $\mathcal{B}$ to violate DDH problem. Thus, the adversary's winning probability in \textbf{Game 2} satisfies the following equation. $$|\Pr[{\sf{Win}}_2]-\Pr[{\sf{Win}}_1]|\leq \epsilon_{DDH}$$

\textbf{Game 3.} Game 3 is the same as Game 2 with one difference. $\mathcal{C}$ replaces the commitment $C_t^*$ with some random number $C_t'$.
Follow the same analysis as in the previous game,
$\mathcal{C}$ sets the public key of $\overline{V_s}$ as $g^a$, $\beta_t$ as $g^b$, $R\in \{g^{ab},g^r\}$, where $r$ is a random number. $\mathcal{C}$ sets $C_t'=g^{v_t}R\cdot{(g^b)}^{\sum_{i=1,i\neq s,t}^nx_i}$.
If there is a difference in the adversary's winning probability between \textbf{Game 2} and \textbf{Game 3}, we can use the adversary to construct an algorithm $\mathcal{B}$ to violate DDH problem. Thus, the adversary's winning probability in \textbf{Game 3} satisfies the following equation. $$|\Pr[{\sf{Win}}_3]-\Pr[{\sf{Win}}_2]|\leq \epsilon_{DDH}$$

\textbf{Game 4.} Game 4 is the same as Game 3 with one difference. $\mathcal{C}$
changes the values of $V_s^*,V_t^*$ with two random elements $V_s'$ and $V_t'$ satisfying certain relation. The change is indistinguishable from $\mathcal{A}$'s view under the DDH assumption.  Wlog, we assume $s<t$. Given the DDH instance $(A = g^a, B = g^b, C)$ where $C\in\{g^{ab},g^r\}$, $\mathcal{C}$ sets the public key of $\overline{V_t}$ and $\overline{V_s}$ as $A=g^a$ and $B=g^b$ respectively.
$\mathcal{C}$ computes $V_s'$ and $V_t'$ as $$V_s'=g^{v_s}A'/C,~~~~~V_t'=g^{v_t}B'C$$ where $$A'={A}^{\sum_{j=1}^{s-1}x_j-\sum_{j=s+1}^{t-1}x_j-\sum_{j=t+1}^{n}x_j},$$ $$B'={B}^{\sum_{j=1}^{s-1}x_j+\sum_{j=s+1}^{t-1}x_j-\sum_{j=t+1}^{n}x_j}.$$ Thus, $V_s'$ and $V_t'$ are two random elements satisfying $V_s'=gA'B'/V_t'$.
Clearly, if there is a non-negligible difference in the adversary's winning probability between \textbf{Game 3} and \textbf{Game 4}, we can use the adversary to construct an algorithm $\mathcal{B}$ to solve DDH problem. Thus, the adversary's winning probability in \textbf{Game 4} satisfies the following equation. $$|\Pr[{\sf{Win}}_4]-\Pr[{\sf{Win}}_3]|\leq \epsilon_{DDH}$$

\textbf{Wrapping up.} The winning probability for $\mathcal{A}$ in \textbf{Game 4} to output a correct \emph{guess} is $1/2$ because in this game the challenges contain only random numbers, which are independent of the votes $v_s,v_t$. Therefore, we can conclude that there is only a negligible difference in winning probability for an adversary between \textbf{Game 0} and \textbf{Game 4}, if all the ZKPoKs in the scheme are {\sf{zero-knowledge}} and DDH assumption holds. So the probability for $\mathcal{A}$ in winning the \emph{MBS} game is $\frac{1}{2}+\epsilon$, where $\epsilon=\epsilon_{ZK}+3\epsilon_{DDH}$.

%
%

\vspace{4 pt}
Now, we show the scheme satisfies fairness, self tallying and dispute freeness as well.

\vspace{4 pt}

\emph{Fairness.} Suppose voter $\overline{V_i}$ votes for $v_i$ in the \emph{Commit} phase and refuse to provide the vote in the \emph{Vote} phase. Due to the {\sf{Soundness}} of ZKPoK, we can guarantee that $v_i$ is decryptable by other voters in the \emph{Recover} phase.



\emph{Self-tallying.} The zero-knowledge proof of knowledge in each algorithm in the proposed protocol forces the voters to perform honestly according to the protocol. After all the voter cast the ballots, the self-tallying property is easy to verify. As $\prod_{i=1}^n{h_i}^{x_i}=0$
in \emph{Tally} algorithm, thus the self-tallying property is achieved.

\emph{Dispute freeness.} For dispute freeness, again the zero-knowledge proof of knowledge in each algorithm of the proposed protocol forces the  commitments and ballots are in the right form and also can be publicly verified by anyone.

\section{Implementation}\label{sec:Implementation}
In this section, we report the implementation results of the proposed construction.

In our experiment, we first implement the protocols on a laptop. For a better simulation of IoT devices, we then implement the protocols on a mobile phone, which has constrained resources. The running environment of the laptop is with Win 8 64-bit operating system and Intel Core (TM) i5-4300 @2.49GHz CPU with an 8 GB SSD. And the configuration of the phone is Android 7.1.1 operating system with Qualcomm MSSM8998 @2.45 CPU (Octa-core) and a 6 GB RAM. The projects are written in C++ language with Miracl library\cite{MIRACL} under Visual Studio 2010 and Android Studio compiler respectively. We test the efficiency of each algorithm when the number of voters increases. The implementation results are illustrated as follows.

\begin{figure}[h]
  \centering
  \includegraphics[width=0.4\textwidth]{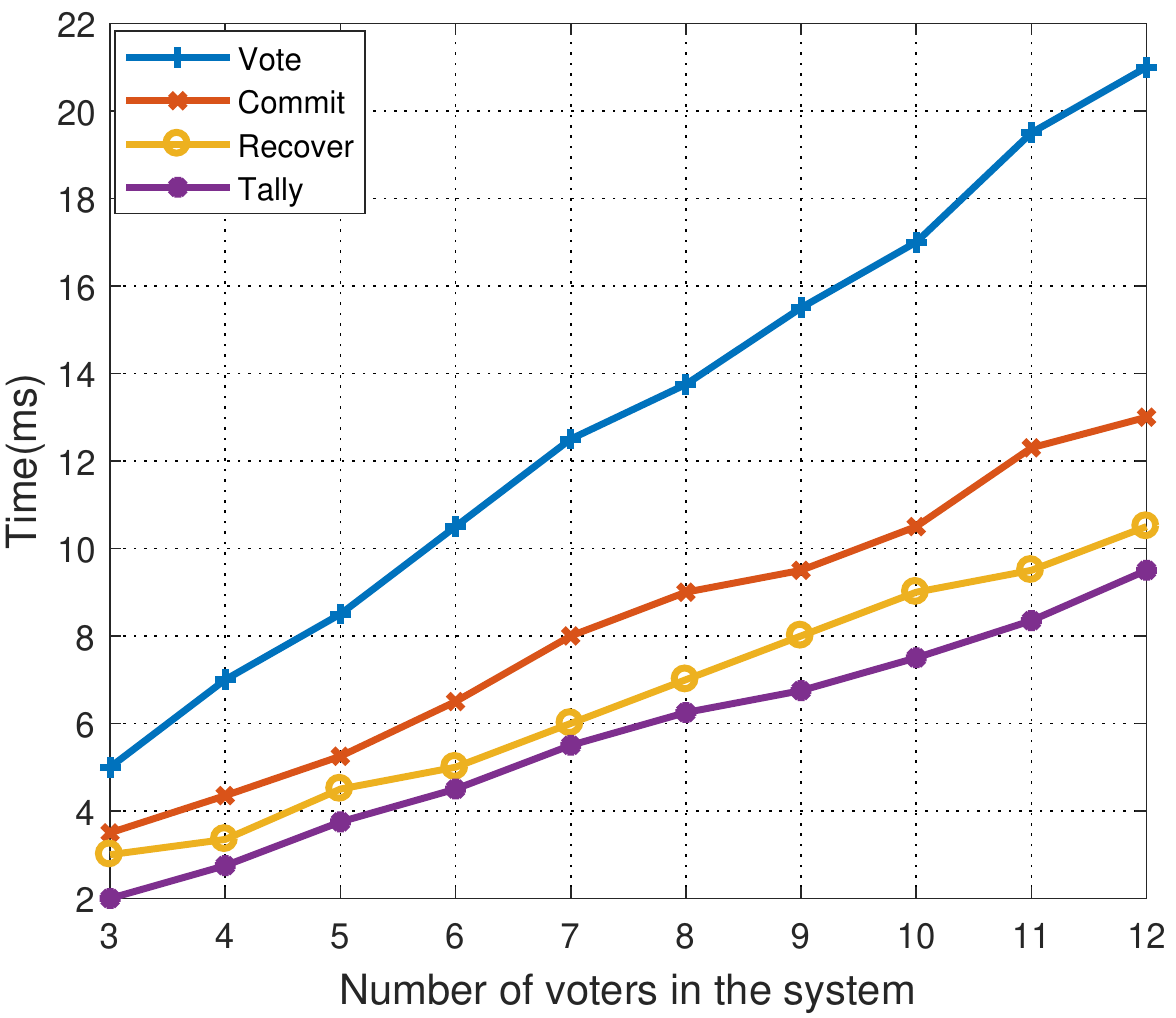}\\
  \caption{Simulation on laptop}\label{test}
\end{figure}

\begin{figure}[h]
  \centering
  \includegraphics[width=0.4\textwidth]{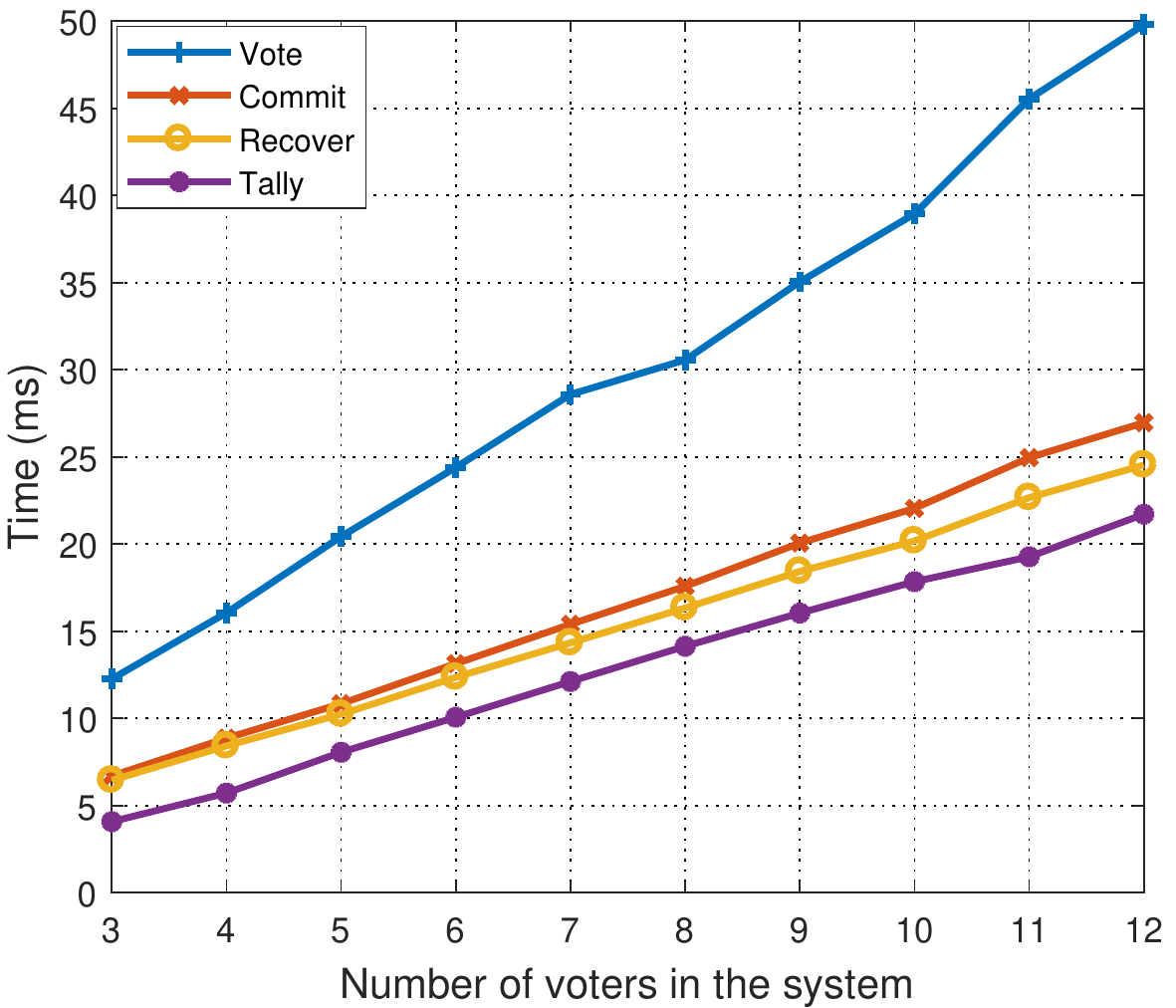}\\
  \caption{Simulation on Android device}\label{testAndroid}
\end{figure}

As we can see from both figures (Fig \ref{test},\ref{testAndroid}) that the running time of each algorithm grows almost linearly with the increase of the number of voters and the two figures have the same trend. Among the four algorithms, the most expensive one is {\sf{Vote}}, as ZKPoK$_3$ is the most complicated one among all the zero-knowledge proofs and it is dominant in {\sf{Vote}}. The running time of {\sf{Vote}} for 3 voters and 12 voters on laptop and phone is 5.01 ms, 12.264 ms and 21.03 ms, 49.794 ms respectively. The fastest algorithm is {\sf{Tally}}, which is also consistent with our empirical analysis, as no zero-knowledge proof is needed and the equation to tally the votes is the product of the voters' ballot, which is linear with the number of voters. The running time of {\sf{Tally}} for 3 voters and 12 voters on laptop and phone is 2.02 ms, 4.076 ms and 9.52 ms, 21.714 ms respectively.  For the other two algorithms, {\sf{Commit}} and {\sf{Recover}}, they grow linearly with the increase of the number of voters, as the computation in these two algorithms requires all the other users' information. Thus, the more voters in the system, the more computation is needed in these two algorithms. For {\sf{Commit}}, the running time on laptop for 3 voters and 12 voters is 3.52 ms and 13.03 ms, and it costs 6.699 ms and 26.963 ms on the phone. For {\sf{Recover}}, the time consumption on laptop for 3 voters and 12 voters is 3.02 ms and 10.51 ms. The time consumption on phone for 3 voters and 12 voters is 6.699 ms and 26.963 ms.

\section{Conclusions}\label{sec:Conclusion}
IoT is pervasive in people's daily life. Decision making is fundamental activity in the IoT. In this paper, we integrate blockchain-based self-tallying voting systems in decentralized IoT architecture. We solve the fairness issues in self-tallying systems with two distinct mechanisms and provide a concrete construction. We prove the security of the construction and also implement it to test the efficiency of the proposed protocol.

 \section*{Acknowledgment}

This work was supported by National Key R\&D Program of China (2017YFB0802000), National Natural Science Foundation of China ( 61872229 ), NSFC Research Fund for International Young Scientists (61750110528), National Cryptography Development Fund during the 13th Five-year Plan Period (MMJJ20170216) and Fundamental Research Funds for the Central Universities(GK201702004).


\vfill\eject

\end{document}